\documentclass[prl,twocolumn,floatfix]{revtex4}
\usepackage[T1]{fontenc} 
\usepackage[cp1250]{inputenc}
\usepackage{graphicx} 
\usepackage[english]{babel} 
\usepackage{color}

\begin{document}

\title{Osmosis with active solutes}

\author{Thomas W. Lion}
\author{Rosalind J. Allen}

\affiliation{SUPA, School of Physics and Astronomy, University of Edinburgh, Mayfield Road, Edinburgh EH9 3JZ, United Kingdom}


\begin{abstract}
Despite much current interest in active matter, little is known about osmosis in active systems. Using molecular dynamics simulations, we investigate how active solutes perturb osmotic steady states.  We find that solute activity increases the osmotic pressure, and  can also expel solvent from the solution - $\emph{i.e.~}$cause reverse osmosis. The latter effect cannot be described by an effective temperature, but can be reproduced by mapping the active solution onto a passive one with the same degree of local structuring as the passive solvent component. Our results provide a basic framework for understanding active osmosis, and suggest that activity-induced structuring of the passive component may play a key role in the physics of active-passive mixtures. 
\end{abstract}

\maketitle

\section{Introduction}

Active particles such as motile bacteria, self-propelled colloids \cite{Golestanian2005,Howse2007,Golestanian2007,Sciortino2009} and vesicles \cite{nardi,thutupalli}, ``chuckers'' \cite{Valeriani2010} and hot nanoparticles  \cite{Radunz2009,Joly2011} are currently of great interest, both from a fundamental perspective and for  their potential applications. Osmotic phenomena play a central role in these systems \cite{Valeriani2010,Brady2008,nardi}, yet osmosis in active systems remains largely unexplored. Here, we use molecular dynamics simulations of a minimal model system to investigate the basic physics of active osmosis. 

Although active systems have been extensively studied with computer simulations, and increasingly with experiments,  few theoretical methods exist for predicting their behaviour. Those that do exist range from the approximate approach of 
 mapping specific  properties of active systems onto those of passive systems
 at a higher ``effective temperature" \cite{Cugliandolo2011,Cugliandolo2008,Joly2011}, to a more recently-derived exact, non-equilibrium, free energy-like relation  for the steady-state collective behaviour of swimming particles
 \cite{Tailleur2008,Cates2013}. It remains unclear, however, how these approaches might translate to osmotic phenomena.

\begin{figure}[h!]
\centering
 {\includegraphics[width=\columnwidth]{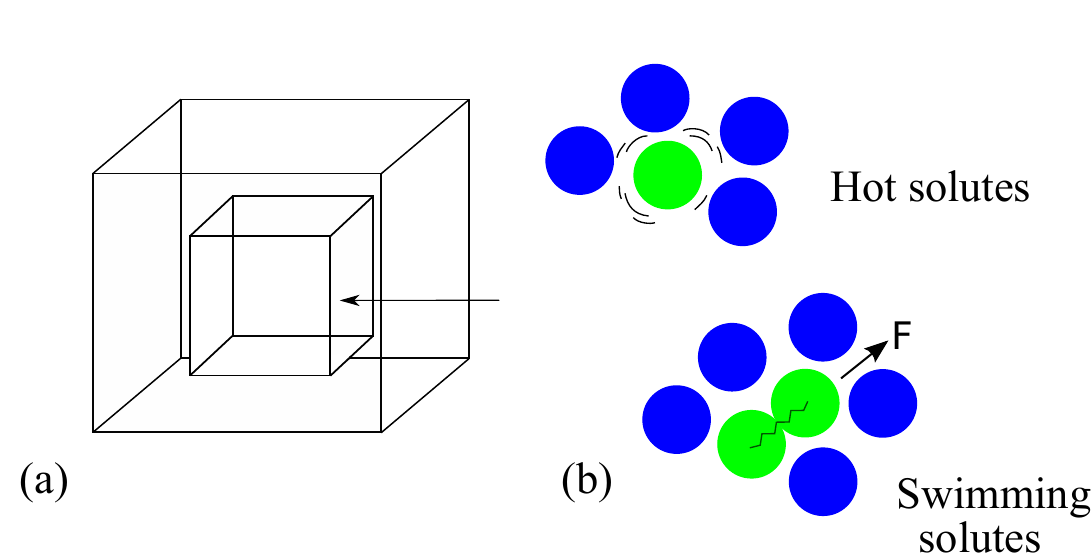}}
 \caption{Our simulation setup. (a) The solution compartment, indicated by the arrow, sits in the centre of the cubic simulation box. (b) Our two active solute models. Solute and solvent particles are shown in green and blue respectively. }
 \label{fig:potential}
\end{figure} 

\begin{figure}[h!]
\centering
 {\includegraphics[width=\columnwidth]{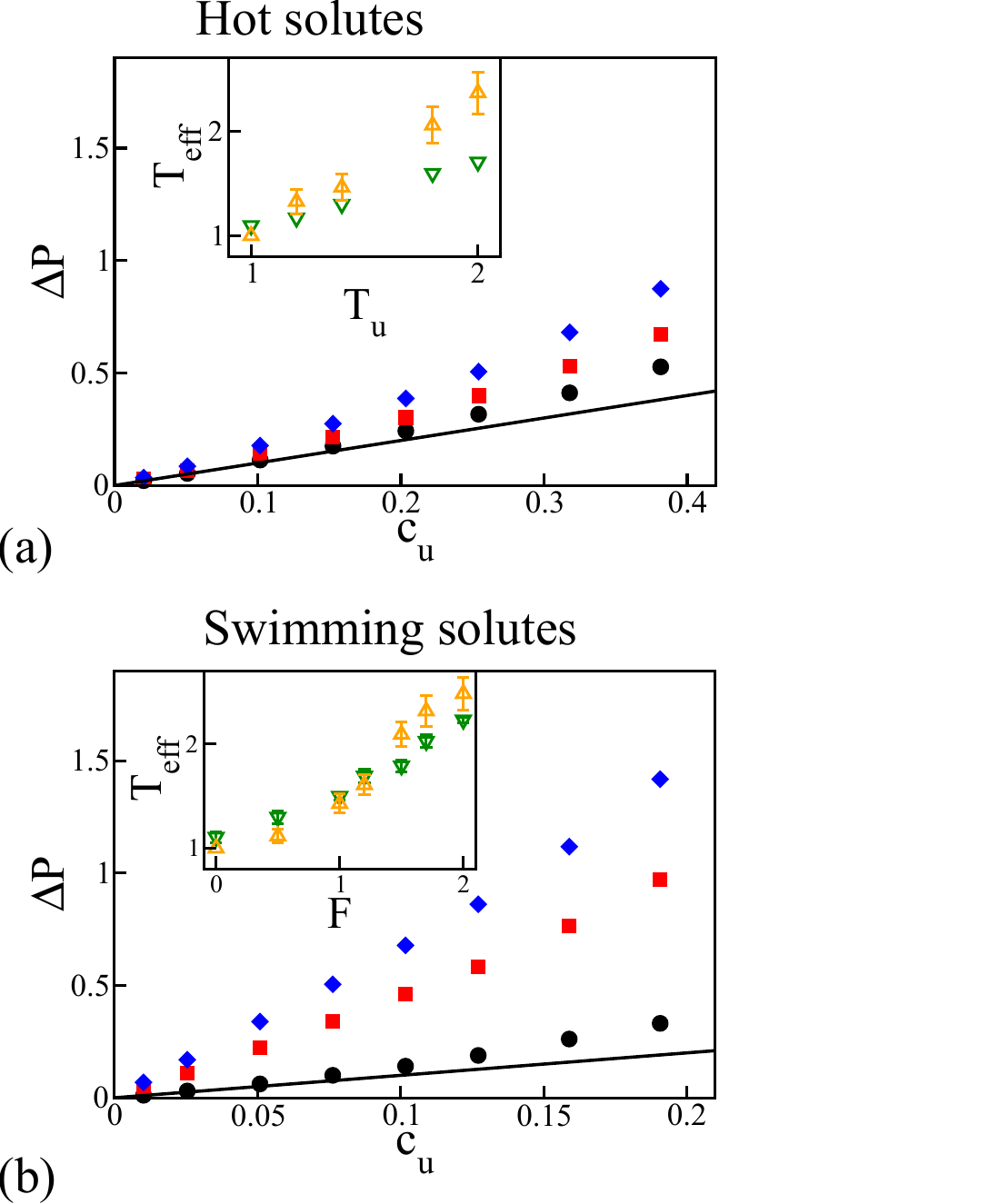}}
 \caption{Active solutes increase the osmotic pressure. The main plots show the osmotic pressure, $\Delta P$, as a function of solute concentration $c_{\mathrm{u}}$, for increasing solute activity. The error bars are smaller than the symbols. In both panels, the black line shows the van 't Hoff relation, Eq.~(\ref{eq:vH}). (a) Hot solutes. Black circles, red squares and blue diamonds represent $T_{\mathrm{u}}=1, 1.4$, and $2.0\,T_{\mathrm{v}}$ (note $k_BT_{\mathrm{v}}=1$). (b)  Swimming solutes. Black circles, red squares and blue diamonds represent $F=0, 4$ and $6$. The insets  show the osmotic effective temperature $T_{\mathrm{osm}}$ (green, down triangles) and the diffusive effective temperature $T_{\mathrm{diff}}$ (orange, up triangles), as functions of the solute activity.  }
 \label{fig:potential2}
\end{figure}

In the classic, passive osmosis experiment, a solution, containing solvent and solute particles, is separated from a pure solvent compartment by a semi-permeable membrane. The solvent distributes itself across the membrane to equalise its chemical potential; at equilibrium  the total density, and the pressure, is higher on the solution side  \cite{vantHoff1887,Gibbs1897}, an effect that can be attributed to the mixing entropy of solvent and solute \cite{Gibbs1897}. From a kinetic point of view, the equilibrium density imbalance can be rationalized by a picture in which a flux of solvent particles out of the solution, driven by the total density gradient, is balanced by an inward diffusive flux \cite{Lion2012eqosmosis}. 

In this paper, we simulate the same experiment, but with active solute particles.  We introduce activity either by maintaining the solutes at an increased temperature (mimicking hot nanoparticles), or by allowing the solutes to be self-propelled (mimicking motile bacteria or active colloids). Because the system is intrinsically out of equilibrium, it is by no means obvious how the solvent will distribute itself across the membrane.  We find that solute activity increases the osmotic pressure, and can qualitatively change the density profiles, such that solvent is driven out of the solution instead of into it, as in passive osmosis \cite{intooutof}. These phenomena cannot be reproduced by an effective temperature, but can be rationalized by a mapping based on the local structure of the solvent. This suggests that local structuring of the passive component may be a key factor controlling the physics of active suspensions.

\section{A minimal model for active osmosis}

 In our simulations, solute and solvent particles are modelled by repulsive spheres with identical interactions, the only difference between them being that the solute particles are confined to a  ``solution compartment'' by a smooth ``membrane potential'', i.e. an external field which is invisible to the solvent \cite{Lion2012pressure,Lion2012eqosmosis} (Fig.~\ref{fig:potential}(a)). The cubic  solution compartment, of length $9.23~\sigma$ (where $\sigma$ is the particle diameter), is located in the centre of a cubic, periodic, simulation box of length $L = 18.42~\sigma$.
Solute and solvent particles interact  via a repulsive, hard-sphere-like Weeks, Chandler, Andersen (WCA) potential \cite{WCA1971}: $U(r) = 4\epsilon((\frac{\sigma}{r})^{12} - (\frac{\sigma}{r})^{6} + \frac{1}{4})$ if $r < 2^{1/6}\sigma$ and zero otherwise; the energy and length scales are defined by setting $\epsilon = k_{B}T_{\mathrm{v}} = 1$ and $\sigma = 1$ (where $T_{\mathrm{v}}$ is the temperature of the passive solvent). The confining potential on the solute particles is of the form  $U = k_{B}T_{\mathrm{v}}(\sigma / \delta)^{9}$, where $\delta$ denotes the perpendicular distance between the particle and the ``membrane''. In total the simulation box contains 5000 particles so that the  total particle density, $\rho_{tot} = 0.8~\sigma^{-3}$ (corresponding to a packing fraction of $0.42~\sigma^{-3}$). The system is simulated using molecular dynamics, with the velocity Verlet algorithm \cite{Allen1992} and a timestep of 0.0005 (in reduced units \cite{Allen1992,redunit}).

Because we are interested in the generic physics of active osmotic systems, we have investigated two very different models for the active solute particles.

{\em{(i) Hot solutes}} The ``hot solutes'' scenario (Fig.~\ref{fig:potential}(b), upper) is inspired by recent work in which a metal nanoparticle performs ``hot Brownian motion'' following the application of laser light \cite{Radunz2009,Joly2011}. In our simulations the solute particles are coupled to a separate thermostat, at a higher temperature  $T_{\mathrm{u}}$ than that of the solvent particles ($T_{\mathrm{v}}$). This induces a net energy flow:  energy is supplied to the solute particles by the solute thermostat, transferred to the solvent particles via solute-solvent collisions, and removed by the solvent thermostat. By using rather strong coupling to the thermostats, it is possible to achieve Maxwell-Boltzmann-like velocity distributions close to the desired temperatures for both sets of particles.
  Here, we use Nos\'{e}-Hoover thermostats \cite{Allen1992} for both solvent and solute; very similar results are obtained using Nos\'{e}-Hoover-Langevin \cite{Leimkuhler2009} and Andersen \cite{Andersen1979} thermostats.  

{\em{(ii) Swimming solutes}} The ``swimming solutes'' scenario (Fig.~\ref{fig:potential}(b), lower) is inspired by  motile bacteria and self-propelled colloids \cite{Howse2007} and vesicles \cite{thutupalli}. In this scenario, solute particles are represented by short dumbbells (pairs of particles connected by a FENE spring with typical bond length $0.96~\sigma$ \cite{Warner1972,fene}) which are propelled by an external force $F$, acting along the long axis of the dumbbell \cite{linek,valeriani2} (the units of $F$ are $k_BT_{\mathrm{v}}/\sigma$). The temperature $T_{\mathrm{v}}$ of the solvent particles is maintained by a  Nos\'{e}-Hoover thermostat (which does not act on the solute dumbbells). In contrast to the hot solutes, the swimming solutes exhibit directional motion, with a correlation length that is limited by  collisions with surrounding particles.

\section{Active solutes increase the osmotic pressure}

 We first investigate how the osmotic pressure is affected by the activity of the solute particles. The osmotic pressure, $\Delta P$, is defined as the  difference in pressure between the solution and solvent compartments. For equilibrium systems,  $\Delta P$ is related to the solute concentration $c_{\mathrm{u}}$, at low solute concentrations, by the van 't Hoff relation \cite{vantHoff1887,Gibbs1897}

\begin{equation}\label{eq:vth}
\Delta P \approx k_{B}Tc_{\mathrm{u}}.
\label{eq:vH}
\end{equation}
Fig.~\ref{fig:potential2} shows that, in our simulations, Eq.~(\ref{eq:vH}) indeed holds in the passive case at low solute concentrations for both our solute models ((a) and (b), black symbols and line). Here, $\Delta P$ is measured by  monitoring the solute-membrane force per unit area; since the solvent does not interact with the membrane, this equals the osmotic pressure difference \cite{Lion2012pressure,Lion2012eqosmosis}. The solution volume (needed to compute the solute concentration) is calculated   by mapping the pressure-density relation of a solute-only simulation onto that of a bulk fluid \cite{Lion2012eqosmosis}. 

In both of our models, increasing the solute activity increases the osmotic pressure (Fig.~\ref{fig:potential2}(a) and (b), red squares and blue diamonds). Interestingly, the relation between $\Delta P$ and $c_{\mathrm{u}}$ remains approximately linear, but with a gradient that increases with solute activity. This suggests that we could define an osmotic effective temperature for the active case, $T_{\mathrm{osm}}$, such that  $\Delta P = k_B T_{\mathrm{osm}}c_{\mathrm{u}}$, at low solute concentrations.  $T_{\mathrm{osm}}$ is shown in the insets to Fig.~\ref{fig:potential2}(a) and (b), and is in quite good agreement with the well-known ``diffusive effective temperature'' $T_{\mathrm{diff}}$, defined by  $T_{\mathrm{diff}} = (D_{\mathrm{u}}/D_{\mathrm{v}})T_{\mathrm{v}}$, where $D_{\mathrm{u}}$ and $D_{\mathrm{v}}$ are the diffusion constants of a single active solute particle in a bath of solvent and  of an equivalent passive particle, respectively \cite{Cugliandolo2008,Joly2011}. However, as we shall see, it turns out that the effective temperature description is not appropriate for describing the active osmotic system, because it fails qualitatively to reproduce the density gradients that are established between the solution and solvent compartments.

\begin{figure}[t!]
 \centering
{\includegraphics[width=\columnwidth]{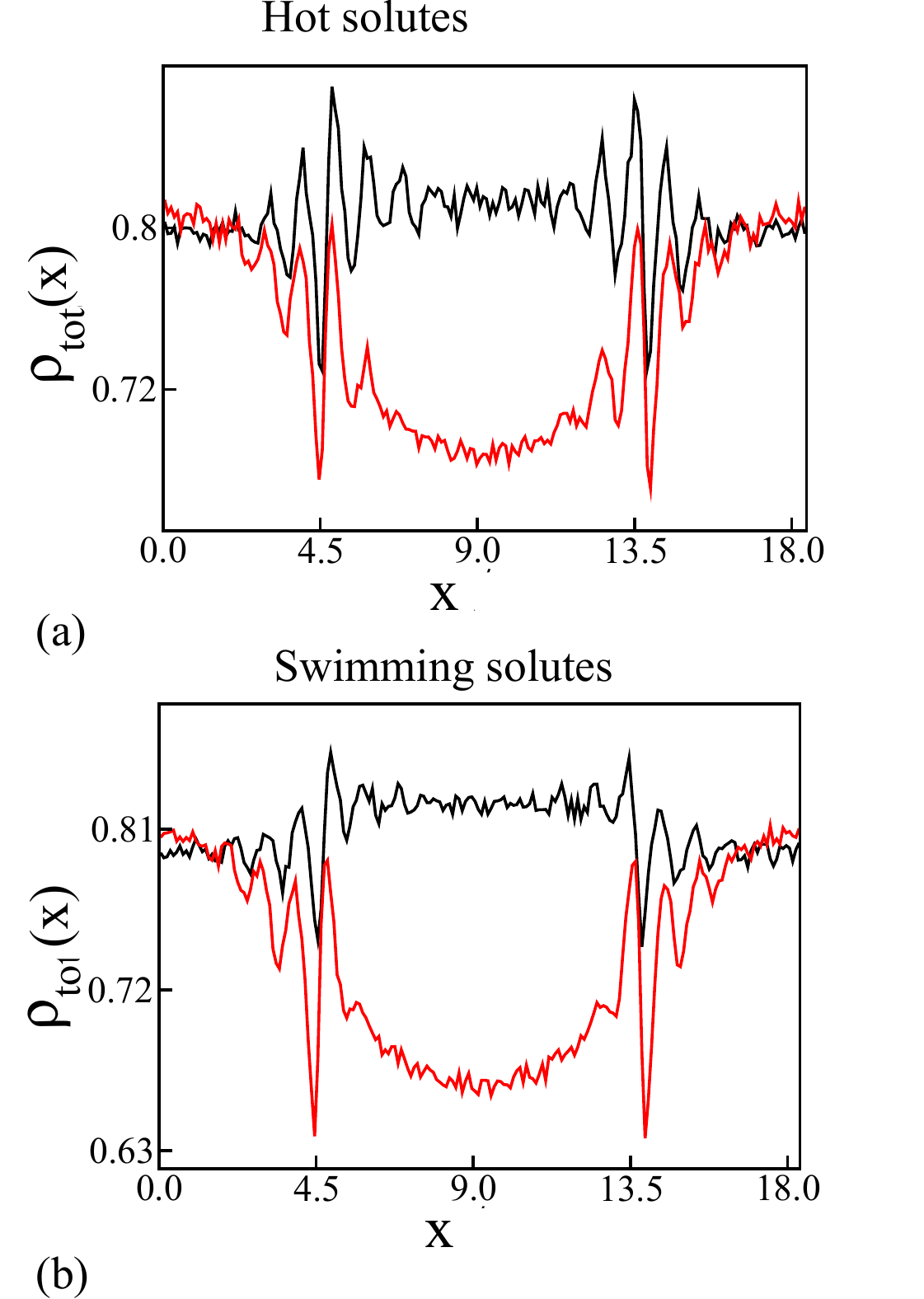}}
 \caption{Steady-state density profiles for passive (black) and active (red) osmotic systems, measured in a slab of width $7.2~\sigma$ through the middle of our simulation box (the units of $x$ are $\sigma$ and the membrane lies at $x= 4.6~\sigma$ and $x=13.8~\sigma$). (a) Hot solutes model with  $c_{\mathrm{u}}= 0.25~\sigma^{-3}$, for $T_{\mathrm{u}} = T_{\mathrm{v}}$ (black) and $T_{\mathrm{u}} = 2T_{\mathrm{v}}$ (red). (b) Swimming solutes model with $c_{\mathrm{u}}= 0.13~\sigma^{-3}$, for $F = 0$ (black) and $F=6$ (red). }
 \label{fig:densityprofile}
\end{figure}

\begin{figure}[t!]
 \centering
{\includegraphics[scale = 0.7,trim=0 0 150 0]{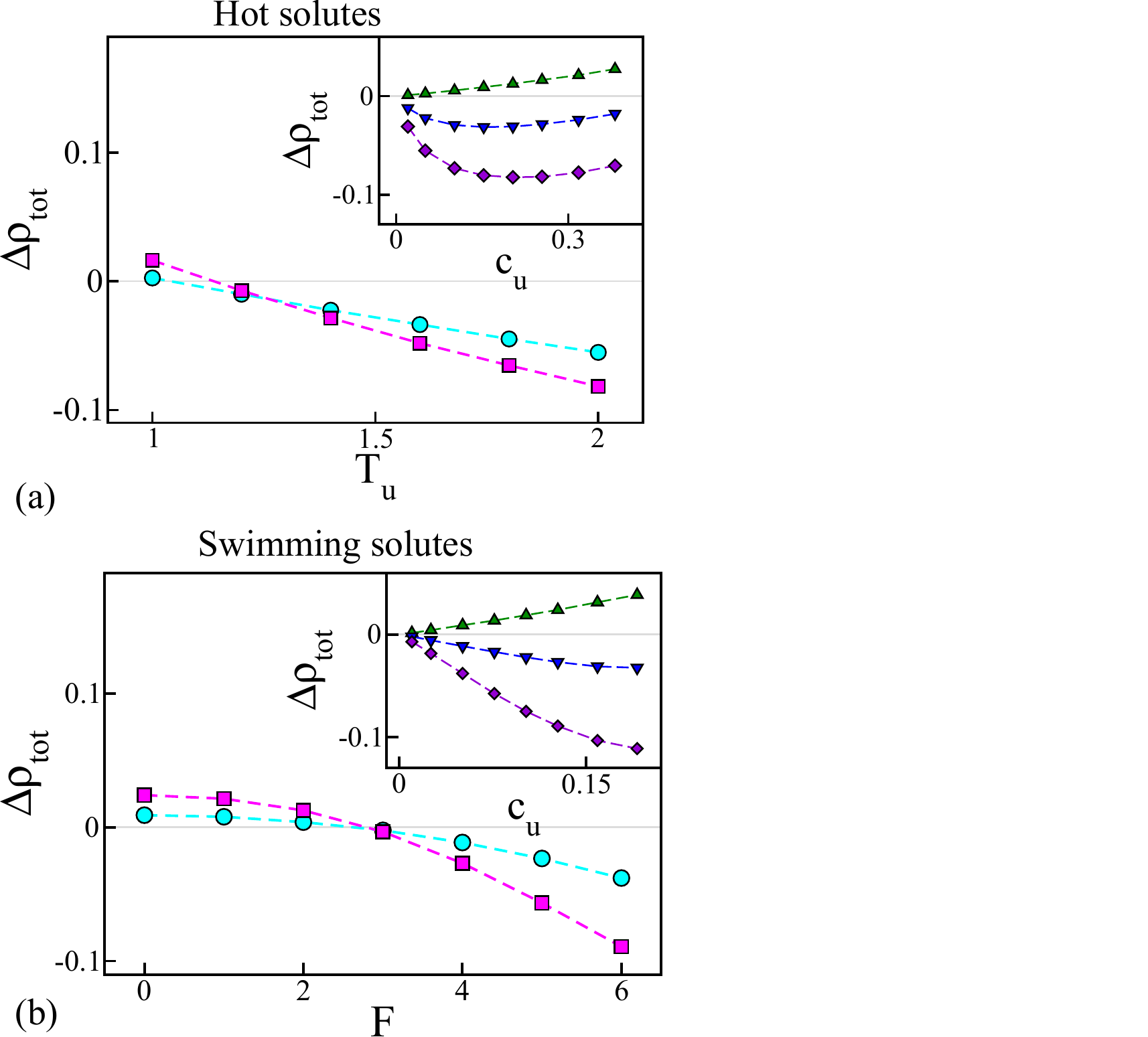}}
 \caption{Reverse osmosis caused by active solutes. The steady-state density difference $\Delta \rho_{tot}$ across the membrane is shown as a function of the solute activity (main plots) and of the solute concentration (insets). Negative values of $\Delta \rho_{tot}$ indicate reverse osmosis. (a) Hot solutes model. Main plot: $\Delta \rho_{tot}$ versus $T_{\mathrm{u}}$ for $c_{\mathrm{u}} = 0.05~\sigma^{-3}$ (cyan circles) and $c_{\mathrm{u}} = 0.25~\sigma^{-3}$ (pink squares). Inset: 
 $\Delta \rho_{tot}$ versus $c_{\mathrm{u}}$ (both in units of $\sigma^{-3}$)  for $T_{\mathrm{u}} = T_{\mathrm{v}}$ (green up triangles), $T_{\mathrm{u}} = 1.4T_{\mathrm{v}}$ (blue down triangles), and $T_{\mathrm{u}} = 2.0T_{\mathrm{v}}$ (purple diamonds). (b) Swimming solutes model. Main plot: $\Delta \rho_{tot}$ versus $F$ for $c_{\mathrm{u}} = 0.05~\sigma^{-3}$ (cyan circles) and $c_{\mathrm{u}} = 0.13~\sigma^{-3}$ (pink squares). Inset: $\Delta \rho_{tot}$ versus $c_{\mathrm{u}}$ for $F = 0$ (green up triangles), $F = 4$ (blue down triangles) and $F=6$ (purple diamonds).}
 \label{fig:densityprofile2}
\end{figure}

\section{Reverse osmosis driven by active solutes}

 For passive systems in osmotic equilibrium, the total particle density is higher on the solution side of the membrane than on the solvent side; this is consistent with the popular picture of osmosis, in which solvent flows from a dilute to a concentrated solution. This behaviour is reproduced in our simulations without activity (Fig.~\ref{fig:densityprofile} (a) and (b), black lines).  
 For active solutes, however, our simulations behave very differently. For  low solute concentrations, the total particle density is {\em{lower}} in the solution than in the solvent compartment (Fig.~\ref{fig:densityprofile} (a) and (b), red lines), indicating that solvent has been driven {\em{out of}} the solution by the solute activity \cite{intooutof}. This picture holds for both our active solute models. Such an effect is known as reverse osmosis; it is usually achieved by applying external pressure to the solution. In our simulations the active solute particles appear to be acting as an ``internal piston'', driving solvent particles out of the solution.  

Fig.~\ref{fig:densityprofile2} shows how the magnitude of the reverse osmotic effect depends on the solute activity and concentration, for the two active solute models. Reverse osmosis corresponds to a negative value of $\Delta \rho_{{tot}}$, the  steady-state difference in total particle density between the solution and solvent compartments. In our simulations, reverse osmosis is only observed above a  critical level of the solute activity (Fig.~\ref{fig:densityprofile2}, main plots). The magnitude of the effect increases with the concentration of the active solutes at low solute concentrations (Fig.~\ref{fig:densityprofile2}, insets), but starts to decrease at  high solute concentrations. This effect is most apparent for the hot solutes at $T_{\mathrm{u}} = 2T_{\mathrm{v}}$. 

Importantly, reverse osmosis driven by solute activity cannot be reconciled with the concept of an osmotic effective temperature. This is because, in a passive osmotic system,  the total steady-state particle density is higher on the solution side of the membrane, regardless of the temperature \cite{Lion2012eqosmosis}; for passive systems, a negative value of  $\Delta \rho_{{tot}}$ can only be produced by applying an external pressure to the solution and not by varying the temperature.

\begin{figure}[t!]
 \centering
 {\includegraphics[width=\columnwidth]{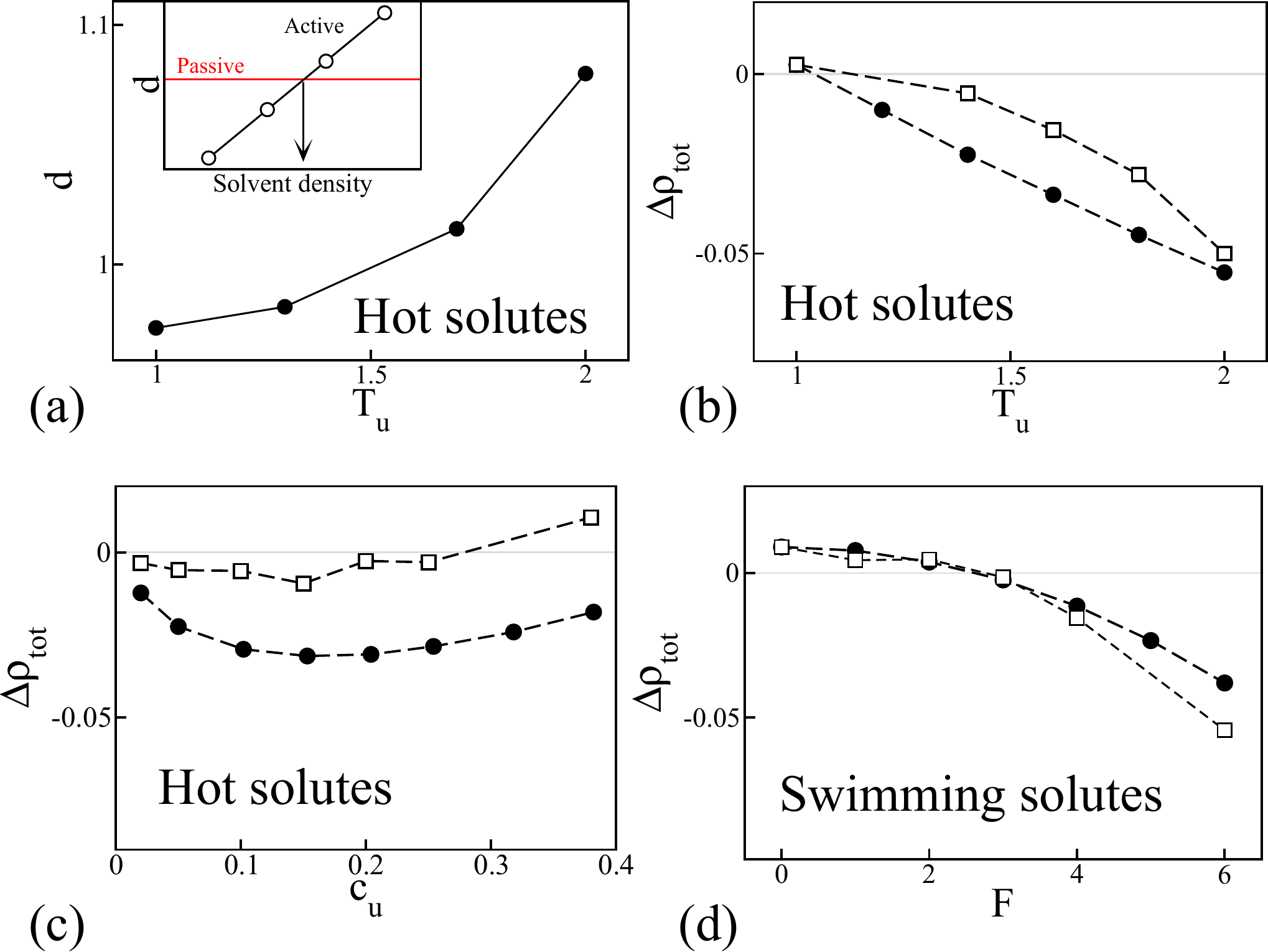}}
\caption{Mapping the solvent local structure onto that of a passive system. (a) Depth $d$ of the first minimum in the PMF (defined as $-k_BT_{\mathrm{v}} \log{g(r)}$) as a function of solute activity $T_{\mathrm{u}}$ for hot solutes at  $c_{\mathrm{u}} = 0.15~\sigma^{-3}$ (for a solution simulated in a periodic box). The inset illustrates our mapping procedure, in which we determine the solvent density for which the value of $d$ in the active solution matches that of a passive osmotic system at the same solute concentration (red horizontal line). (b)-(d) Comparison between the prediction of the matching procedure (open symbols) for the total osmotic density imbalance $\Delta \rho_{{tot}}$  and the results of our simulations (filled symbols). (b) Hot solutes model: $\Delta \rho_{{tot}}$ as a function of $T_{\mathrm{u}}$, for $c_{\mathrm{u}} = 0.05~\sigma^{-3}$. (c) Hot solutes model:  $\Delta \rho_{{tot}}$ as a function of  $c_{\mathrm{u}}$, for   $T_{\mathrm{u}} = 1.4 T_{\mathrm{v}}$. (d) Swimming solutes model:  $\Delta \rho_{{tot}}$ as a function of $F$, for $c_{\mathrm{u}} = 0.05~\sigma^{-3}$.}
 \label{fig:grplots}
\end{figure}

\section{What causes active solute-driven reverse osmosis?}

 We speculate that reverse osmosis occurs in our simulations because the active solutes act as an ``internal piston'', pushing the solvent particles away from them and ultimately forcing them out of the solution. In other words, active solute particles exclude more volume from occupation by solvents than passive solutes would do. Supporting this view, we observe that the  solvent particles experience a more structured local environment as the solute activity increases, suggesting that they are being pushed together by the activity of the solutes. As a measure of the local structure close to a solvent particle, we measure the depth $d$ of the first minimum of the potential of mean force, defined as $-k_BT_{\mathrm{v}}\log{g(r)}$ \cite{JPHansen} (where  the computation of the radial distribution function $g(r)$ counts all particles at distance $r$ from a given solvent particle). Fig.~\ref{fig:grplots}(a) shows that $d$ increases as the solute activity $T_{\mathrm{u}}$ increases in the hot solutes model, for a fixed concentration of solutes; equivalent results are obtained for swimming solutes. A similar effect on the radial distribution function has recently been observed for  passive colloids in a bath of motile bacteria \cite{Angelani2011}.

For passive systems, the solvent's chemical potential on either side of the membrane is directly related to its density; equalising the solvent chemical potential across the membrane and taking into account the contribution of the solute density allows one to derive the van `t Hoff relation \cite{Lion2012eqosmosis}. In our  systems, the active  solution is intrinsically out of equilibrium, so that the chemical potential is only defined for the solvent particles outside the solution region. We hypothesize, however, that one might be able to define a ``chemical potential-like'' quantity for solvent particles inside the active solution region, analogous to the passive chemical potential, but using a {\em{local}} effective density rather than the global solvent density. This is equivalent to postulating that the solvent particles behave as though they were in a passive system, but experiencing a higher local density in the solution region, due to the presence of the active particles.  

To test this hypothesis, we attempted to map the behaviour of our active osmotic systems onto equivalent passive systems, by matching the value of the local solvent structuring (as measured by $d$). To this end,  we measured how  the value of $d$ varied with solvent density for an active solution at fixed solute concentration $c_{\mathrm{u}}$. This allowed us to determine the solvent density for which the value of $d$ in the active solution matched that of an  equivalent passive solution (Fig.~\ref{fig:grplots}(a), inset) - i.e. the solvent density for which the degree of solvent structuring in the osmotic solution matched that of the passive solution at the same solute concentration. Assuming that this would be the steady-state solvent density in the active osmotic solution, we could make a prediction for  the osmotic density imbalance $\Delta \rho_{tot}$ for the active osmotic systems.

 Fig.~\ref{fig:grplots}(b)-(d)  show the results of this procedure, for the two active solute models, for different solute activities and concentrations.  The matching procedure indeed predicts reverse osmosis, above a threshold activity level, and with magnitude increasing with solute activity, in reasonable quantitative agreement with our simulations (Fig.~\ref{fig:grplots}(b) and (d)).  The procedure also correctly reproduces the decrease in magnitude of the reverse osmotic effect for high solute concentrations (Fig.~\ref{fig:grplots}(c)) \cite{cluster}. Thus, matching the local solvent structure between active and passive systems does seem to provide a way to understand the underlying (non-equilibrium) physics that is involved. Further work will be needed to determine whether the concept of an effective chemical potential for the passive component of an active-passive mixture, defined on the basis of local structure, has wider utility.


\section{Discussion}

 In this paper, we have investigated the basic physics of active osmotic systems, using a minimal model system in which the solute and solvent particles are repulsive and interact identically and the semi-permeable membrane is represented by a smooth confining potential acting only on the solutes. Our simulations show that solute activity increases the osmotic pressure, and can cause reverse osmosis. The fact that similar results are obtained for two very different active solute models suggests that these phenomena may be rather generic. Although we have focused here only on systems with identical solute and solvent particle size and interparticle interactions, we note that, for passive systems, osmotic phenomena are remarkable for their generality (the   van `t Hoff law applies regardless of the nature of the solute and solvent particles) -- suggesting that similar universality may hold for these non-equilibrium active systems.

From a practical point of view, reverse osmosis caused by an externally applied pressure is widely used in water purification. Do our results imply that similar results could be achieved using active particles? Probably not, for several reasons. Perhaps most obviously, in suspensions of  motile microorganisms, Janus particles or laser-heated nanoparticles, the active particles are typically at much lower concentration than in our simulations, so that the reverse osmotic effect would probably be very small. It is also important to note that, for the hot solutes model, we have assumed a very strong coupling to the active thermostat. In real systems of hot solutes, such as laser-heated nanoparticles, heat dissipation to the solvent is likely to severely limit the levels of solute activity that can be achieved. Our results should, however, be  relevant to colloidal systems containing mixtures of active and passive  repulsive particles. Here the passive particles would play the role of the solvent and optical tweezers  \cite{Angelani2011}, ``electric bottles''  \cite{Leunissen2008-1} or  suitable microfluidic devices \cite{galajda} could be used for selective confinement. Suspensions of active colloids are currently attracting much attention for their novel phase behavior \cite{theurkauff,Wensink2012,farrell,cates,redner,stenhammar,stenhammar2,Menzel2013}, and active-passive mixtures should show even richer physics \cite{mccandlish}. Osmotic effects are very likely to play a key role in the behavior of such systems. 

Interestingly, we have observed that our system can be mapped onto an equilibrium one, by matching a simple measure of the local structuring of the solvent particles. This suggests a new way of using equilibrium thermodynamic concepts to describe active-passive systems. Our work suggests that local structuring of the passive component, caused by the active particles, may be central to understanding the physics of these systems. The connections between this idea and both the ``active stress'' that is used to account for non-equilibrium driving in the hydrodynamic studies of active fluids  \cite{simha}, and the active pressure component that has recently been found to emerge from a continuum description of active phase separation \cite{stenhammar,wittkowski}, remain to be elucidated (although we note that comparison of the osmotic pressure between active and passive systems apparently provides a direct way to measure the active stress). Clarifying these links, as well as the connection to active particle phase separation recently seen in purely active suspensions, should lead to a deeper understanding of the connection between equilibrium and active systems in general.

From a practical point of view, we also note that an osmotic simulation setup of the type used here provides a convenient way to probe numerically the physics of active-passive mixtures. The chemical potential of the passive component is well-defined  in the region outside the active solution compartment, and solvent particles are free to exchange between compartments. Thus, our setup allows for ``semi-grand canonical'' simulation of active-passive mixtures, even though, from a theoretical point of view, the chemical potential inside the solution compartment is not defined.

\acknowledgments
We thank Mike Cates, Davide Marenduzzo, Wilson Poon and Joakim Stenhammar for helpful discussions. TWL was supported by an EPSRC DTA studentship and RJA by a Royal Society University Research Fellowship. This work was performed using the Edinburgh Compute and Data Facility, which is partially supported by the eDIKT initiative.


\begin{thebibliography}{44}
\bibitem[{int()}]{intooutof}
\bibinfo{note}{Here we define ''out of'' and ''into'' relative to a starting
  state with uniform density on each side of the membrane.}
\bibitem[{red()}]{redunit}
\bibinfo{note}{The reduced unit of time is $m\sigma^{2}/\epsilon$, where $m$ is
  the mass of a particle, here set to unity.}
\bibitem[{fen()}]{fene}
\bibinfo{note}{$U_{FENE} = U_{WCA} -
  \frac{\kappa}{2}R_0^2\log\left[1-\left(\frac{r}{R_0}\right)^2\right]$ if $r <
  R_0$ and $\infty$ otherwise. Here, $\kappa =30$ denotes the spring constant
  and $R_0=1.5\,\sigma$ is the maximum extension of the spring.}
\bibitem[{clu()}]{cluster}
\bibinfo{note}{We speculate that the decrease in magnitude for high solute
  concentrations may be due to the break-up of solvent clusters by more
  frequent collisions with the active solutes at higher solute concentrations.}
\end{thebibliography}

\end{document}